\documentclass[prc,aps,amsmath,amssymb,nofootinbib,superscriptaddress,12pt]{revtex4-2}
\usepackage{epsfig}
\usepackage[bookmarksnumbered,bookmarksopen,colorlinks,citecolor=blue,linkcolor=blue]{hyperref}

\begin{document}

    \title{ Systematic study of capture thresholds with time dependent Hartree-Fock theory }

    \author{Hong Yao}
    \affiliation{ School of Physics, Beihang University, Beijing 102206, People's Republic of China}
    \affiliation{ Department of Physics, Guangxi Normal University, Guilin 541004, People's Republic of
        China }

    \author{Hui Yang}
    \affiliation{ Department of Physics, Guangxi Normal University, Guilin 541004, People's Republic of
        China }

    \author{Ning Wang}
    \email{wangning@gxnu.edu.cn}\affiliation{ Department of Physics,
        Guangxi Normal University, Guilin 541004, People's Republic of
        China }
    \affiliation{ Guangxi Key Laboratory of Nuclear Physics and Technology, Guilin 541004, People's Republic of
        China }

    \begin{abstract}
        With the time dependent Hartree-Fock (TDHF) theory, capture thresholds $E_{\rm cap}$ for 144 fusion systems with nearly spherical nuclei are systematically studied for the first time. We find that for the reactions between doubly-magic nuclei, the calculated $E_{\rm cap}$ are very close to the extracted barrier heights from measured fusion excitation functions. For the fusion reactions with nearly spherical nuclei, an excitation energy of about 1 MeV at the capture position need to be considered to better reproduce the data due to the lower excitation threshold. The rms deviation with respect to the barrier heights is only 1.43 MeV from the TDHF calcualtions, which is smaller than the results from three empirical nuclear potentials. Together with Siwek-Wilczy\'{n}ski formula in which the three parameters are determined by the TDHF calculations, the measured fusion cross sections at energies around the barriers can be well reproduced for seven fusion reactions $^{40}$Ca+$^{48}$Ca, $^{16}$O+$^{208}$Pb, $^{40}$Ca+$^{90,96}$Zr, $^{28}$Si+$^{96}$Zr and $^{132}$Sn+$^{40,48}$Ca.
    \end{abstract}

     \maketitle

\newpage
    \begin{center}
        \textbf{I. INTRODUCTION}
    \end{center}

    The heavy-ion fusion reaction at energies around the Coulomb barrier is an important way not only for the synthesis of  super-heavy nuclei (SHN) \cite{Hof00,Ogan10,Sob,Koz16,Mori20,Ren02,Pomo18,Adam04,Zhu16,Hind96,Hind21} and extremely neutron-deficient nuclei \cite{Zhang19,Zhang21,Zhou21}, but also for the study of the nuclear structures \cite{Stok78, Sarg11, Das98,Jia14}. For light and intermediate fusion systems, the fusion (capture) cross sections can be described by using the fusion coupled channel calculations \cite{Hag99,Dasso87,Das98} or empirical barrier distribution approaches \cite{Zag01,SW04,liumin,Wang09,Wangbing,Jiang22} together with some static nuclear potentials \cite{Wong73,Bass74,Bass80,Tian07,BW91,Gup92,Wen22}. For fusion systems leading to the synthesis of SHN, the quasi-fission (QF) wherein the composite system fails to evolve into a compound nucleus (CN) after capture and breaks apart before reaching compact equilibrium shapes, significantly complicates the description of fusion process, and is difficult to be described by using the traditional barrier penetration approach. To understand the dynamical process in fusion and to investigate the influence of dynamical effects, shell effects and isospin effects on the fusion barrier, some microscopic dynamics models such as the TDHF \cite{Naka05,Maru06,Guo07,Sim12,Sim14,Stev16} and the improved quantum molecular dynamics (ImQMD) model \cite{ImQMD2010,ImQMD2014} have been used, with which the neck dynamics and the contact time of the composite system can be self-consistently described.

    In recent years, the TDHF theory is successfully used in the study of the fusion barriers, especially in the collisions between two spherical nuclei. Based on the density-constrained TDHF (DC-TDHF) method \cite{Umar85}, which combines TDHF dynamics with a minimization technique under constraints on the one-body density, the energy-dependent nucleus-nucleus potential can be precisely obtained. For light and medium mass systems such as $^{40}$Ca+$^{40}$Ca and $^{16}$O+$^{208}$Pb, DC-TDHF calculations at near barrier-top energy give a fusion barrier which is expected to match the TDHF fusion threshold. It is found that the fusion thresholds from the TDHF calculations are in good agreement with the experimental data \cite{Wash08,Umar14,Sim13}. Up to now, the fusion cross sections for more than a thousand of reaction systems have been measured in the past several decades. A systematic comparison between the experimental data and the TDHF predictions especially for the collisions between open shell nuclei, is therefore very interesting and necessary.

    Very recently, the fusion barriers of more than 300 fusion systems are extracted through fitting the measured fusion/fission  excitation functions with a modified Siwek-Wilczy\'{n}ski fusion cross section formula \cite{Chen23}. In this work, we would like to systematically investigate the capture thresholds for the 144 reaction systems with nearly spherical nuclei (with charge number of nuclei $Z\geqslant 6$ and the energies of the lowest excited state of reaction partners being larger than 1 MeV) listed in Ref. \cite{Chen23} by using the TDHF theory. Here, the capture threshold $E_{\rm cap}$ is defined as the energy above which central collisions lead to capture while the exit channel at lower energies is made of two outgoing fragments (via quasi-elastic or deep inelastic scattering). For light and intermediate fusion systems, the capture barrier is almost the same as the fusion barrier. However, for heavy systems, such as the reactions leading to SHN, the fusion barrier could be higher than the capture barrier since an extra push is needed to achieve fusion \cite{Swiat82}. To systematically investigate the capture thresholds for different reaction systems, one needs to discriminate capture from fusion for heavy systems.

    It is known that the reaction time in deep inelastic scattering is shorter than that in QF \cite{Toke82}. Therefore, it could be possible using the contact time of the composite system to discriminate capture from fusion for heavy systems. It is thought that QF takes place within a few zeptoseconds ( $1 zs = 10^{-21}s$) \cite{Toke82,Shen87,Hind92,Riet11,Riet13}, faster than that of fission of the CN. For example, the average contact times extracted from the measured QF mass-angle distributions vary from about 3 zs for $^{64}$Ni+ $^{238}$U to about 5 zs for $^{50}$Ti+ $^{249}$Cf, at energies extending from below to about 10\% above the capture barriers \cite{Albe20}, and the mean asymmetric QF time for $^{86}$Kr+$^{198}$Pt and $^{86}$Kr+$^{197}$Au was found to be about 3 zs in Ref. \cite{Sen22}. On the other hand, it is found that the relaxation time for the radial energy dissipation in $^{208}$Pb induced deep inelastic reactions at incident energies close to the Coulomb barrier is of the order of 1 zs \cite{Rehm81}. In addition, the contact times of the composite systems in head-ion collisions of both $^{58}$Ni+$^{124}$Sn \cite{Wu19} and $^{48}$Ca+$^{208}$Pb \cite{Sun22} at energies just below the capture barriers are shorter than 1.6 zs from the TDHF calculations. The ImQMD calculations for $^{58}$Ni+$^{208}$Pb at energies 10\% above the Coulomb barriers also indicate that the quasi-elastic and the deep-inelastic collisions occur when the contact time is less than about 1.4 zs \cite{Li19}.

    It is therefore reasonable to set a contact time of about 2 zs ($\sim$ 600 fm/c) as the time of capture for the composite system. If the composite system reseparates into two fragments within 2 zs after projectile-target contact, we treat it as a scattering process rather than QF in this work.

    The structure of this paper is as follows: In Sec. II, the frameworks of TDHF will be introduced. In Sec. III, the capture thresholds $E_{\rm cap}$ for 144 fusion systems will be systematically calculated with TDHF, at the same time, the capture barriers and capture excitation functions will be further studied based on the obtained $E_{\rm cap}$. Finally a summary is given in Sec. IV.

    \begin{center}
        \textbf{ II. THEORETICAL FRAMEWORKS AND DETAILS IN CALCULATIONS }\\
    \end{center}

    In the TDHF theory, the complicated many-body problem is replaced by
    an independent particle problem, i.e., the many-body wave functions are approximated as the anti-symmetrized
    independent particle states to assure an exact
    treatment of Pauli principle during time evolution. In the nuclear context, the basic ingredient of TDHF is
    the energy functional composed by the various one-body densities. Here, we adopt the Skyrme energy density functional with the parameter set SLy6 \cite{SLy6}.   The dynamical evolution of the mean-field is expressed by TDHF equation
    \begin{equation}
        i \hbar \frac{d \hat{\rho}}{dt} =[ \hat{h}[\hat{\rho}],\hat{\rho}],
    \end{equation}
    with the single-particle Hamiltonian $h[\hat\rho]$ and the one-body density $\rho$.
    Taking the nuclear ground state as an initial state of the dynamical evolution, TDHF time
    evolution is determined by the dynamical unitary propagator.
    Earlier TDHF calculations imposed the various approximations
    on the effective interaction and geometric symmetry. The development of computational power
    allows a fully three-dimensional (3D) TDHF calculation with the
    modern effective interaction and without symmetry restrictions, which significantly improves the physical scenario in heavy-ion collisions \cite{Guolu14}. In this work, the code Sky3D is used \cite{Maru14}.

    In the dynamic TDHF evolution, the coordinate-space grid consists of $n_x \times n_y \times n_z$ points which increase with the size of nuclear system. We set
    \begin{eqnarray}
        n_x=n_y= \left\{
        \begin{array} {r@{\quad:\quad}l}
            24   &   Z\leq30    \\
            30   &   30<Z\leq60    \\
            40   &   60<Z\leq100    \\
        \end{array} \right.
    \end{eqnarray}
    for different nuclear systems, with grid spacing of 1 fm and the charge number $Z$ of the heavier nuclei in reactions. $n_z\approx 5(R_1+R_2)$ with the size $R_1=1.4A_1^{1/3}$ and $R_2=1.4A_2^{1/3}$. Here, $A_1$ and $A_2$ denote the mass number of the projectile and target nuclei, respectively.  The time propagation is carried out using a Talyor-seriers expansion up to the sixth order of the unitary mean-field propagator with a time step of 0.2 fm/c. The initial distance of two nuclei is set as $2(R_1+R_2)$ fm and the impact parameter is set as zero (head-on collision) for the systematic calculations of the capture thresholds.

    After projectile-target contact, we let the composite system further evolves 2 zs to see whether it reseparates into two fragments or not, with which one knows the reaction process is capture or scattering. To ascertain the contact between the projectile and the target, we employ the DBSCAN (Density-Based Spatial Clustering of Applications with Noise) algorithm \cite{Est96} to identify the number of clusters during the simulation process, which is similar to the cluster recognition algorithm in the ImQMD model \cite{ImQMD20}. In the $x-z$ plane, the sample set is composed of grid points with a density exceeding half of the saturation density, i.e., $\rho \geqslant 0.08  \, {\rm fm}^{-3}$.
    %The neighborhood value is determined by $\sqrt{dx^2 + dz^2}$, and the minimum point count is set to 1.

    \begin{center}
        \textbf{III. RESULTS AND ANALYSIS}
    \end{center}

    In this section, we first introduce the calculation of capture threshold $E_{\rm cap}$.
    Then, the capture barrier will be analyzed by considering the excitation energy of the composite system at capture position.
    Finally, the capture threshold is further applied on the calculation of fusion (capture) cross sections.

    \begin{center}
        \textbf{A. Capture Thresholds}
    \end{center}

    \begin{figure}
        \setlength{\abovecaptionskip}{-1.4cm}
        \includegraphics[angle=0,width=0.75\textwidth]{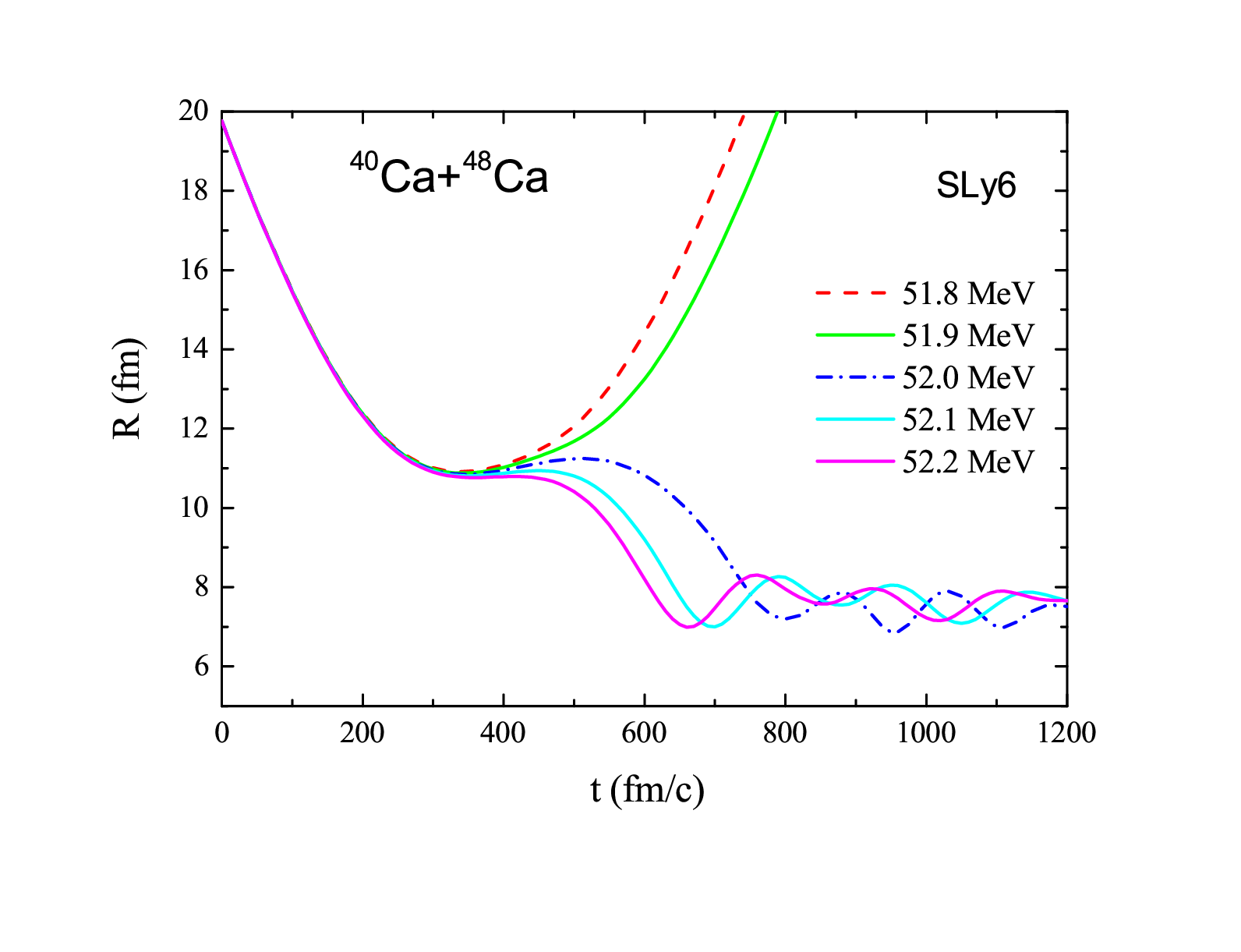}
        \caption{(Color online) Time evolution of the center-to-center distance between the fragments in $^{40}$Ca + $^{48}$Ca. }
    \end{figure}

    In Fig. 1, we show the time evolution of the distance between the fragments in $^{40}$Ca + $^{48}$Ca at central collisions. One can see that at an incident energy of 51.9 MeV, the center-to-center distance rapidly increases after $t=400$ fm/c  due to the Coulomb repulsion, which indicates a process of scattering occurs. Whereas, with a slight increase of incident energy by 0.1 MeV, the distance between the two fragments falls to about 7.5 fm at $t=1000$ fm/c  which indicates the capture occurs. The capture threshold of $51.95\pm 0.05 $ MeV can be unambiguously obtained for $^{40}$Ca + $^{48}$Ca.

    \begin{figure}
        \setlength{\abovecaptionskip}{-0.6cm}
        \includegraphics[angle=0,width=0.75\textwidth]{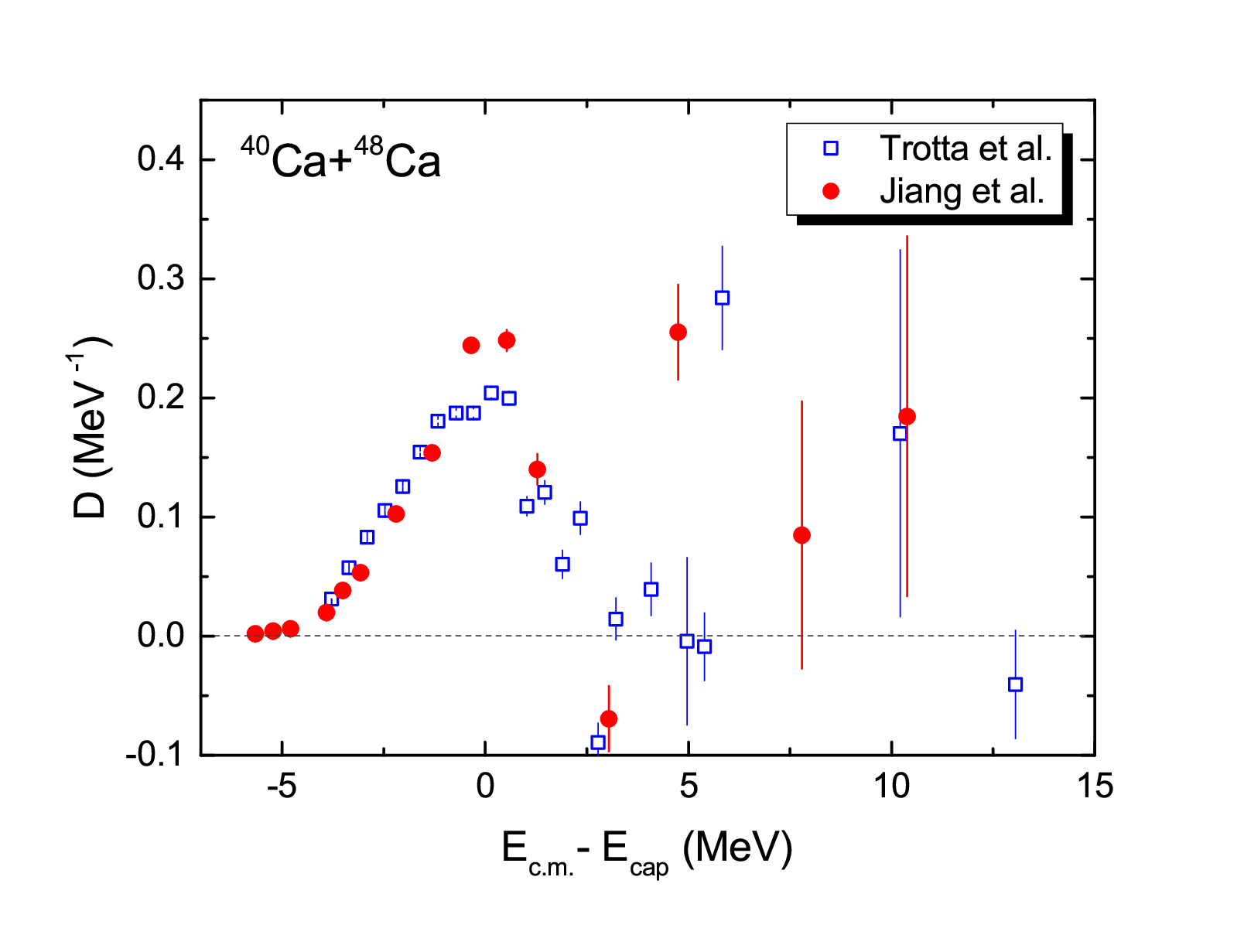}
        \caption{(Color online) Experimental barrier distribution for $^{40}$Ca + $^{48}$Ca. The horizontal axis is shifted by the TDHF capture threshold of $E_{\rm cap}=51.95$ MeV.  }
    \end{figure}

    In Fig. 2, we show the extracted barrier distribution for $^{40}$Ca + $^{48}$Ca from the measured fusion excitation functions \cite{Trot01,Jiang10}, by using $D(E)=\frac{1}{\pi R_0^2} \frac{d^2 (E \sigma)}{d E^2}$. The values of barrier radius $R_0$ are taken from \cite{Chen23}. The horizontal axis is shifted by the TDHF capture threshold $E_{\rm cap}=51.95$ MeV.  One can see that the TDHF capture threshold is very close to the peak position of the barrier distribution.

    \begin{figure}
        \setlength{\abovecaptionskip}{-0.7cm}
        \includegraphics[angle=0,width=0.85\textwidth]{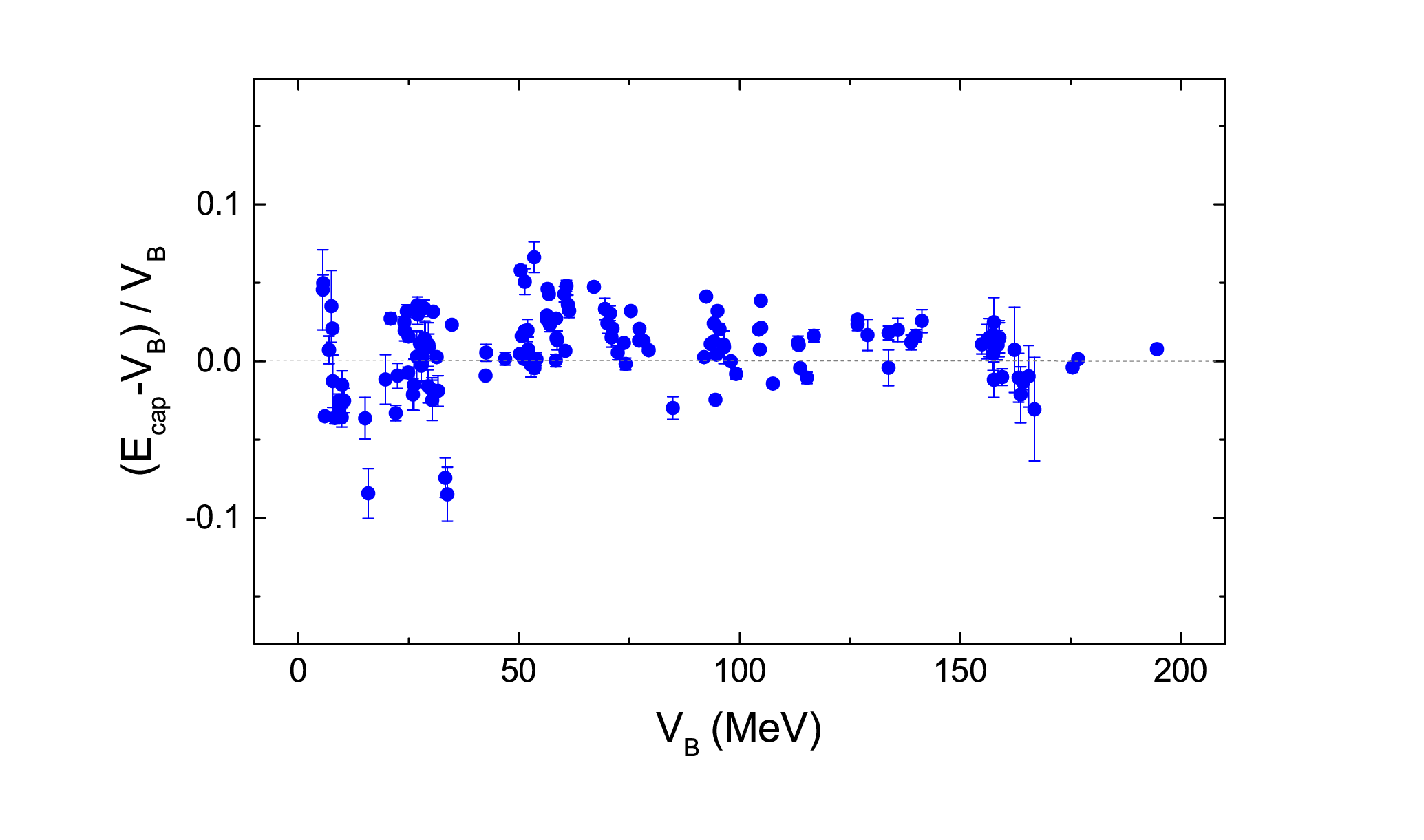}
        \caption{(Color online) Relative deviation between the TDHF capture threshold and the extracted barrier height $V_B$ \cite{Chen23}. The error bars denote the uncertainties of the extracted barrier heights in the fit to the measured cross sections. }
    \end{figure}

    For fusion reactions induced by nuclei with small deformations, the orientations of nuclei in the entrance channel could influence the nucleus-nucleus interaction potential and consequently the capture threshold. For reactions with spherical nuclei bombarding on deformed target, there are two extreme orientations in the reaction plane leading to two extreme barrier heights. From the TDHF calculations for $^{16}$O + $^{154}$Sm and $^{16}$O + $^{238}$U, one notes that the fusion barriers at side collisions are more close to the peak positions of the experimental barrier distributions \cite{Sim08}. For reactions with two deformed nuclei, e.g. oblate projectile bombarding on prolate target nucleus, there exists four extreme orientations in the reaction plane. In this work, we investigate the dynamics process at four extreme orientation configurations, i.e., side-side, tip-tip, tip-side and side-tip collisions. We then obtain four thresholds with $E^{(1)}_{\rm cap} \geqslant E^{(2)}_{\rm cap} \geqslant E^{(3)}_{\rm cap} \geqslant E^{(4)}_{\rm cap}$ for a certain reaction.  For a given heavy fusion system with deformed nuclei, the TDHF calculations require long CPU times \cite{God19}, since one needs to perform TDHF run at different $E_{\rm c.m.}$ energy for a fixed impact parameter and orientation angle, and the calculations of the distribution of the capture thresholds are therefore computationally too demanding. To give a systematic comparison of the TDHF calculations, we empirically write the (most probable) capture threshold as,
    \begin{equation}
        E_{\rm cap} \simeq \frac{1}{3}(E^{(1)}_{\rm cap} + E^{(2)}_{\rm cap})+\frac{1}{6}(E^{(3)}_{\rm cap} + E^{(4)}_{\rm cap})
    \end{equation}
    for fusion reactions with nearly spherical nuclei, considering that the configurations with relatively higher thresholds ($E^{(1)}_{\rm cap}$ and $E^{(2)}_{\rm cap}$) corresponding to those with larger geometry radii (i.e., prolate nuclei at side collision or oblate at tip one) are more probable \cite{Sim08}. We note that the calculations with Eq.(3) are comparable with the most probable barrier height from the CCFULL calculations in which the contributions from six individual orientation angles are analyzed \cite{Hagi12}. As a test, we calculate the mean value of the capture thresholds for $^{28}$Si + $^{96}$Zr with the integral over impact parameter $b$ and orientation angle of $^{28}$Si assuming an isotropic distribution of the orientations at a fixed impact parameter. The contributions of the centrifugal potentials are removed in the calculations of the capture thresholds. The obtained mean value of the capture thresholds for collisions with $b\leqslant 4$ fm is 71.32 MeV, which is very close to the result of $E_{\rm cap}=71.69$ MeV according to Eq.(3), with a relative deviation of $0.5\%$. It indicates that the approximation in Eq.(3) is reasonable.

    In the systematic calculations of the TDHF capture thresholds, we use the binary search method which is terminated when the energy interval is smaller than 0.2 MeV. Fig. 3 shows the relative deviation between the TDHF capture threshold and the extracted barrier height $V_B$ \cite{Chen23} for the 144 fusion reactions with nearly spherical nuclei. One can see that the relative deviations (with an rms error of $2.48\%$) are smaller than $5\%$ for most of reactions, which also implies
    the weights in Eq.(3) for different orientation configurations are reasonable.

\vspace{3em}
    \begin{center}
        \textbf{B. Capture Barriers}
    \end{center}

    According to the energy conservation, one has \cite{ImQMD2010}
    \begin{equation}
        E_{\rm c.m.}=T+V+E^{*},
    \end{equation}
    for head-on collisions. Where $E_{\rm c.m.}$ is the incident center-of-mass energy, $T$ is
    the relative motion kinetic energy of two fragments and $E^*$ is the excitation energy.
    At the closest approach with an incident energy of the capture threshold $E_{\rm cap}$, the kinetic energy $T=0$. The height of the capture barrier $V_B$ is therefore expressed as,
    \begin{equation}
        V_B =E_{\rm cap}- E_{\rm cap}^{*}.
    \end{equation}
    Where, $E_{\rm cap}^{*}$ denotes the (most probable) excitation energy of the composite system at the capture position. Before projectile-target contact, the excitation energy of the system could be negligible for spherical nuclei, $V_B \approx E_{\rm cap}$ is therefore expected as shown in Fig. 3.

    To investigate the influence of $E_{\rm cap}^{*}$, we introduce a quantity: excitation threshold $\varepsilon_{th}$, which is defined as the energy of the lowest excited state of the reaction partners \cite{E1E2}. The values of $\varepsilon_{th}$ for reactions with two spherical nuclei are generally larger than those with deformed nuclei. One finds that the energies of the first excited states for neutron shell-closed nuclei are generally larger than one MeV \cite{Long19}. In this work, we focus on the 144 fusion reactions with nuclei of $\varepsilon_{th}>1$ MeV. If the excitation energy of the composite system at capture position is lower than the excitation threshold, we assume the two nuclei remain ground states at the moment and consequently $V_B = E_{\rm cap}$ for most of colliding events.

    Considering that the excitation energy of the reaction system is very small before contact of the reaction partners \cite{Umar09} and its upper limit is the sum of $E_{\rm c.m.}$ and the reaction $Q$-value, we write the excitation energy at capture position as
    \begin{equation}
        E_{\rm cap}^{*} \approx c \, (E_{\rm cap}+Q)- \varepsilon_{th},
    \end{equation}
    with the coefficient $c=0.052$. The data of $\varepsilon_{th}$ are taken from \cite{E1E2}.
    %The lower the value of $\varepsilon_{th}$ is, the higher the value of $E_{\rm cap}^{*}$ would be for a certain reaction system.
    For the cases with $c \, (E_{\rm cap}+Q)<\varepsilon_{th}$ which usually occur in reactions with doubly-magic nuclei, we set $E_{\rm cap}^{*}=0$ as mentioned previously.

    \begin{table}
        \caption{ The rms deviation (in MeV) between model predictions and the extracted barrier heights \cite{Chen23} for the 144 reaction systems with nearly spherical nuclei ($\varepsilon_{th} >1$ MeV). }
        \begin{tabular}{ccccc }
            \hline
            & ~TDHF~  &  ~Bass\cite{Bass74}~  & ~BW91\cite{BW91} ~ & ~MWS\cite{Wang08}~   \\
            \hline
            & 1.43	   & 2.11                  &  1.67	            &	1.53	           \\

            \hline
        \end{tabular}
    \end{table}

    \begin{figure}
        \setlength{\abovecaptionskip}{-0.5cm}
        \includegraphics[angle=0,width=0.85\textwidth]{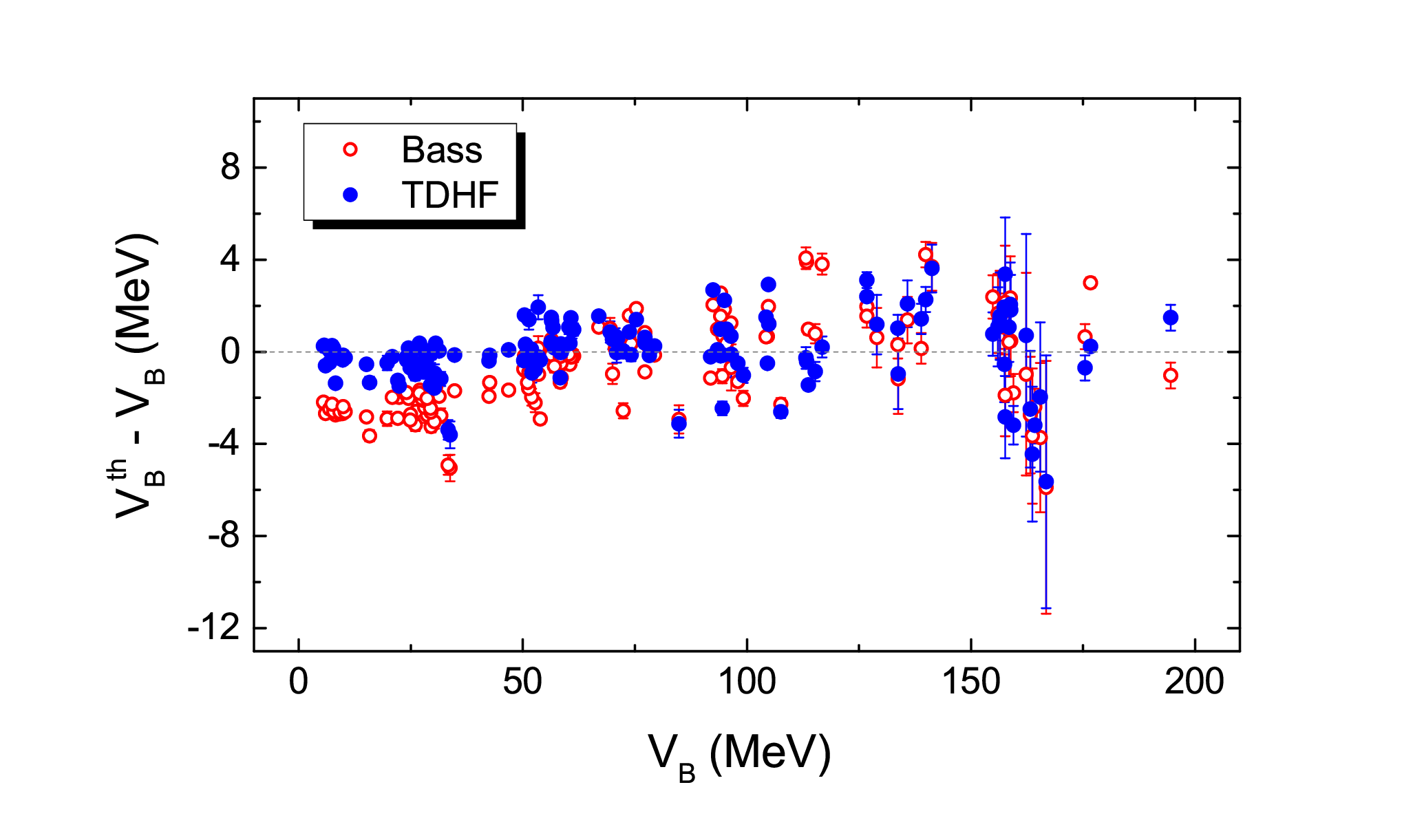}
        \caption{(Color online) Deviations between the calculated barrier heights and the extracted data \cite{Chen23}. The solid circles and the open ones denote the results from the TDHF calculations and those from the Bass potential \cite{Bass74}, respectively. The error bars denote the uncertainties of the extracted barrier heights. }
    \end{figure}

    The capture barriers of fusion reactions can be roughly estimated by using some parameterized formulas which are expressed as functions of charge and mass number of projectile and target, such as the Bass potential \cite{Bass74},
    \begin{equation}
        V_B^{\rm Bass}=\frac{Z_1 Z_2 e^2}{1.07 \, (A_1^{1/3}+A_2^{1/3})+2.70}-2.90 \frac{A_1^{1/3} A_2^{1/3}}{A_1^{1/3}+A_2^{1/3}}.
    \end{equation}
    In Table I, we list the root-mean-square (rms) deviation between the predictions from four models and the extracted  barrier heights for a total of 144 reaction systems. Here, both BW91 \cite{BW91} and MWS \cite{Wang08} adopt Woods-Saxon form of nuclear potential, but with different model parameters. In MWS, the model parameters are obtained from the Skyrme energy density functional (with the parameter set SkM*\cite{Bart82}) combining the extended Thomas-Fermi approach and the frozen density approximation. Through introducing an empirical barrier distribution with a superposition of two Gaussian
    (2G) functions \cite{liumin,Wang09}, the most probable barrier height can be obtained $V_B^{\rm MWS}\approx 0.946 B_0$, with the barrier height $B_0$ of the Woods-Saxon potential. One sees that the rms deviation from the TDHF calculations is only 1.43 MeV, which is smaller than the results of the three empirical nuclear potentials. We also note that the rms deviation is reduced by $16\%$ considering the influence of $E_{\rm cap}^{*}$ in the TDHF calculations. In Fig. 4, we show the discrepancy between the calculated barrier heights and the extracted ones. The solid circles denote the results from the TDHF calculations according to Eq.(5). Here the results (open circles) of the Bass potential \cite{Bass74} are also presented for comparison. One sees that for light systems, the results from the TDHF calculations are much better. For some heavy fusion systems such as $^{58,64}$Ni+$^{112-124}$Sn, the large deviations with both TDHF and Bass potential could be due to the large uncertainties of the extracted barrier heights in the fit to the measured cross sections. From the relative errors in Fig. 3 and the absolute errors in Fig. 4, one can analyze the model accuracy for describing light systems and that for heavy ones, considering that the barrier height changes from a few MeV for light reaction systems to hundreds MeV for heavy ones. In addition, with the absolute errors, one knows the difference between the model accuracy for describing barrier heights and that for nuclear masses.

    \begin{center}
        \textbf{C. Capture cross sections}
    \end{center}

    In Ref. \cite{SW04}, Siwek-Wilczy\'{n}ska and Wilczy\'{n}ski (SW) propose an analytical fusion cross section formula, with a single-Gaussian distribution of barrier heights for describing the barrier distribution,
    \begin{equation}
        \sigma_{\rm {fus}}=\pi R_B^{2}  \frac{W}{\sqrt{2}E_{\rm c.m.}}
        [X  {\rm erfc}(-X)+\frac{1}{\sqrt{\pi}}\exp(-X^{2}) ],
    \end{equation}
    where $X = (E_{\rm c.m.}-V_B)/\sqrt{2}W $. $V_B$ and $W$ denote the centroid and the standard deviation of the Gaussian function, respectively.  $R_B$ denotes the barrier radius. In this work, we use the SW formula for describing the capture cross section together with the three parameters being determined from the TDHF calculations. Here, $V_B$ is determined by the capture threshold according to Eq.(5). The barrier radius $R_B$ is from the frozen Hartree-Fock (FHF) calculations based on the same parameter set SLy6. The standard deviation of the Gaussian function $W$ mainly relates to the dissipation effect and the deformation effect.

    For fusion reactions with doubly-magic nuclei, such as $^{16}$O+$^{208}$Pb, the authors in Ref. \cite{Sim08} conclude that the TDHF calculations contain dynamical effects that reduce the barrier by $\sim$ 2 MeV as compared to the frozen approximation. In addition, the barrier distribution is smeared out with a finite width of $ {\rm FWHM} \approx 0.56 \hbar \omega$, typically $2\sim3$ MeV, in the quantum mechanical treatment of a single parabolic potential barrier, which includes tunnelling \cite{Das98}. Considering the finite width of the barrier distribution, we set $W=2$ MeV for the fusion reactions with stable doubly-magic nuclei.

    For the fusion reactions with nearly spherical nuclei, such as $^{40}$Ca+$^{90, 96}$Zr, we set $W=c(E_{\rm cap}+Q)$ to effectively consider the dissipation effect which is related to the excitation energy of the system at the capture position. According to Eq.(5), the barrier height that the reaction partners "feel" is influenced by the excitation energy since a part of effective incident energies are transformed into the inner excitations of the system rather than to overcome the potential barrier. A relatively higher excitation energy at capture position could result in stronger effects for dynamical deformations and nucleon transfer in the capture process at sub-barrier energies and broadens the width of the barrier distribution. Wolski also note that the $Q$-value effect should be taken into account when comparing various fusion data \cite{Wol13}.
    %Here, the reduction factor $(1-c)$ is to consider the fact that the excitation threshold for nearly spherical nucleus is generally larger that of strongly deformed nucleus, and a relatively lower effective excitation energy at the capture position is therefore expected comparing with the deformed one.

    \begin{figure}
        \setlength{\abovecaptionskip}{-2.2cm}
        \includegraphics[angle=0,width=0.85\textwidth]{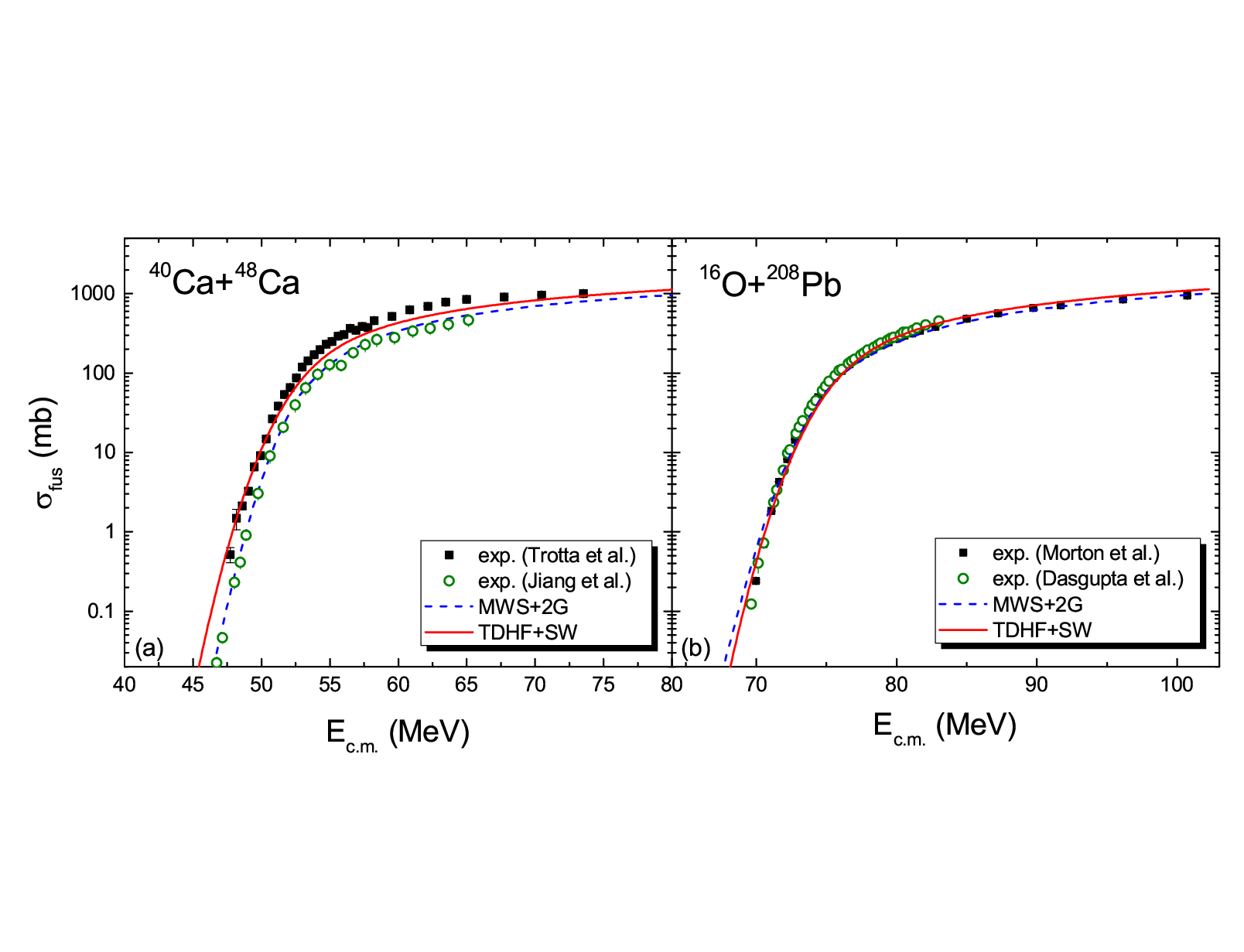}
        \caption{(Color online) Fusion excitation functions for $^{40}$Ca+$^{48}$Ca and $^{16}$O+$^{208}$Pb. The experimental data in (a) are taken from \cite{Trot01} and \cite{Jiang10}. The data in (b) are taken from \cite{Mort99} and \cite{Das97}.}
    \end{figure}

    \begin{table}
        \caption{ TDHF calculated parameters in SW formula. The reaction $Q$-value and the most probable barrier height $V_B^{\rm MWS}$ based on the modified Woods-Saxon potential \cite{Wang08} are also listed.}
        \begin{tabular}{cccccc}
            \hline\hline

            ~~~reaction~~~  & ~~~$V_B$ (MeV)~~~ & ~~~W (MeV)~~~ & ~~~$R_B$ (fm)~~~ & ~~~$Q$ (MeV)~~~ & ~~~$V_B^{\rm MWS}$ (MeV)~~~ \\
            \hline
            $^{40}$Ca+$^{48}$Ca  &  51.98   &  2      &  10.2   &	 $4.56$   &  52.43    \\
            $^{16}$O+$^{208}$Pb  &  74.65   &  2      &  11.6   &  $-46.48$  &  74.43 	\\
            $^{40}$Ca+$^{90}$Zr  &  97.08   &  2.10   &  10.7   &  $-57.02$  &  97.85    \\
            $^{40}$Ca+$^{96}$Zr  &  95.07   &  2.87   &  11.0   &	$-41.09$  &  97.14    \\
            $^{28}$Si+$^{96}$Zr  &  70.55   &  2.73   &  10.3   &  $-19.28$  &  69.82    \\
            $^{132}$Sn+$^{40}$Ca &  114.38  &  3.24   &  11.5   &  $-52.13$  &  115.71   \\
            $^{132}$Sn+$^{48}$Ca &  112.66  &  1.92   &  11.8   &  $-75.78$  &  112.73   \\

            \hline\hline
        \end{tabular}
    \end{table}

    We list in Table II the parameters in the SW formula obtained with the TDHF calculations for seven fusion reactions. The relative deviations between the barrier heights from the TDHF calculations and those from the MWS potential are generally smaller than $2\%$. In Fig. 5, we show the fusion excitation functions of $^{40}$Ca+$^{48}$Ca and $^{16}$O+$^{208}$Pb. The solid curves denote the results of TDHF calculations together with the SW formula. The dashed curves denote the results of empirical barrier distribution approach \cite{Wang08,Wang09} in which a superposition of two Gaussian (2G) functions is used for describing the barrier distribution together with the modified Woods-Saxon (MWS) potential to determine the barrier parameters. One can see that the data can be reproduced reasonably well by the model predictions, considering the uncertainty of the measured cross sections.
    We note that with the Pauli exclusion principle between nucleons being taken into account in the FHF calculations \cite{Sim17}, the barrier radii are not changed significantly in comparison with the usual FHF potentials. We also note that the barrier radii from the FHF calculations with SLy6 are 10.2 fm and 11.6 fm for $^{40}$Ca+$^{48}$Ca and $^{16}$O+$^{208}$Pb, respectively, which are systematically larger than those of the MWS potential by about 0.6 fm. The predicted fusion cross sections with SLy6 at above barrier energies are consequently larger than those of MWS due to the larger geometry radii.

    \begin{figure}
        \setlength{\abovecaptionskip}{-2.2cm}
        \includegraphics[angle=0,width=0.85\textwidth]{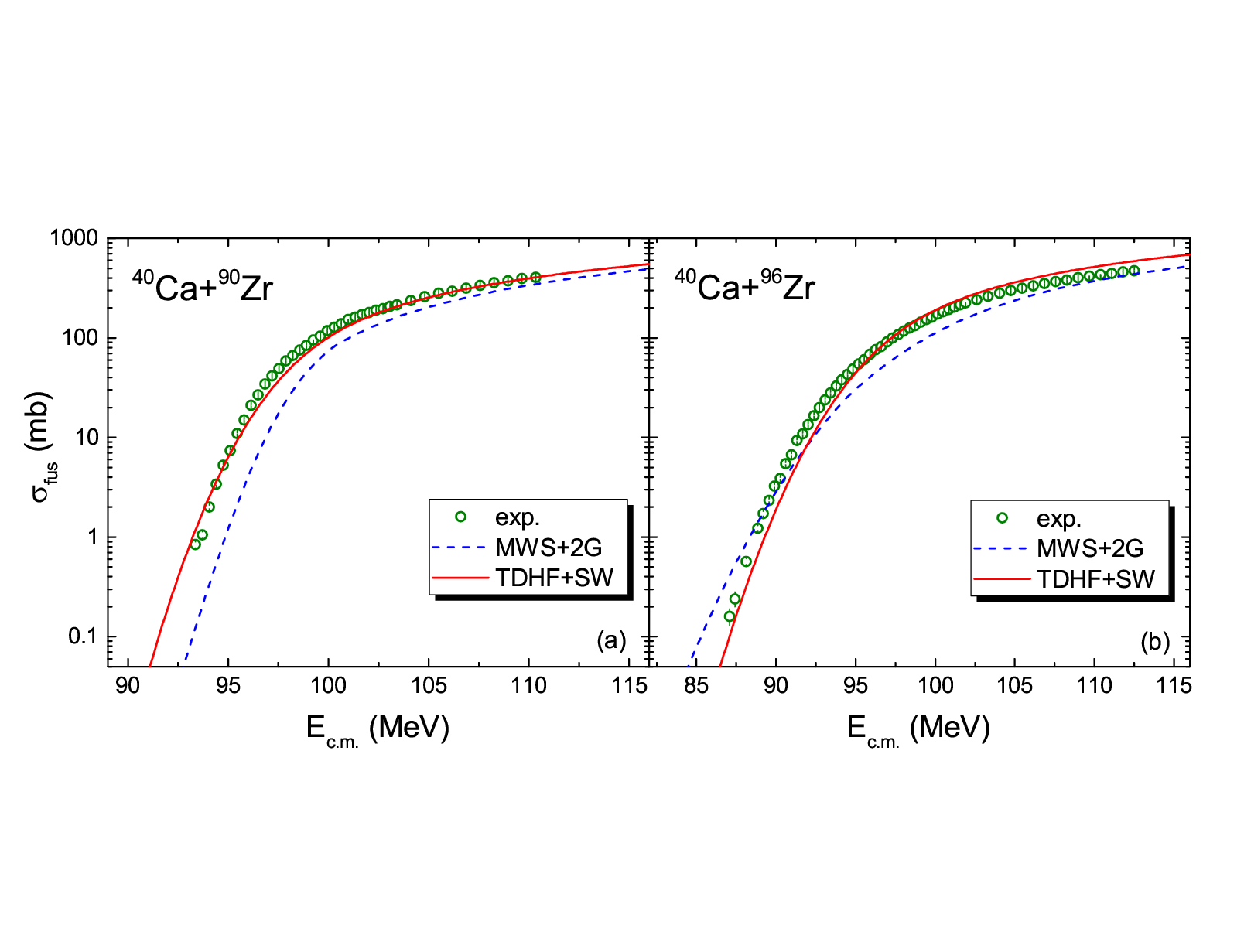}
        \caption{(Color online) The same as Fig. 5, but for $^{40}$Ca+$^{90,96}$Zr. The data are taken from \cite{Timm98}.}
    \end{figure}

    Fig. 6 shows the fusion excitation functions of $^{40}$Ca+$^{90,96}$Zr. We note that the reaction $Q$-value for $^{40}$Ca+$^{96}$Zr is larger than that for $^{40}$Ca+$^{90}$Zr by 16 MeV, which could result in a relatively larger dissipation due to the larger excitation energy of the system in the neutron-rich nuclei induced reaction $^{40}$Ca+$^{96}$Zr. From Table II, one sees that the value of $W=2.87$ MeV for $^{40}$Ca+$^{96}$Zr is obviously larger than that for $^{40}$Ca+$^{90}$Zr.  With the capture barriers from the TDHF calculations, the experimental fusion cross sections for the both of the reactions can be well reproduced. We also note that the results from the TDHF calculations are better than those from the MWS+2G approach. Because the most probable barrier height from the MWS+2G approach is higher than the extracted one \cite{Chen23} by about 1.5 MeV for $^{40}$Ca+$^{90}$Zr, the corresponding fusion cross sections calculated are obviously lower than the experimental data at sub-barrier energies.

    \begin{figure}
        \setlength{\abovecaptionskip}{-0.3cm}
        \includegraphics[angle=0,width=0.65\textwidth]{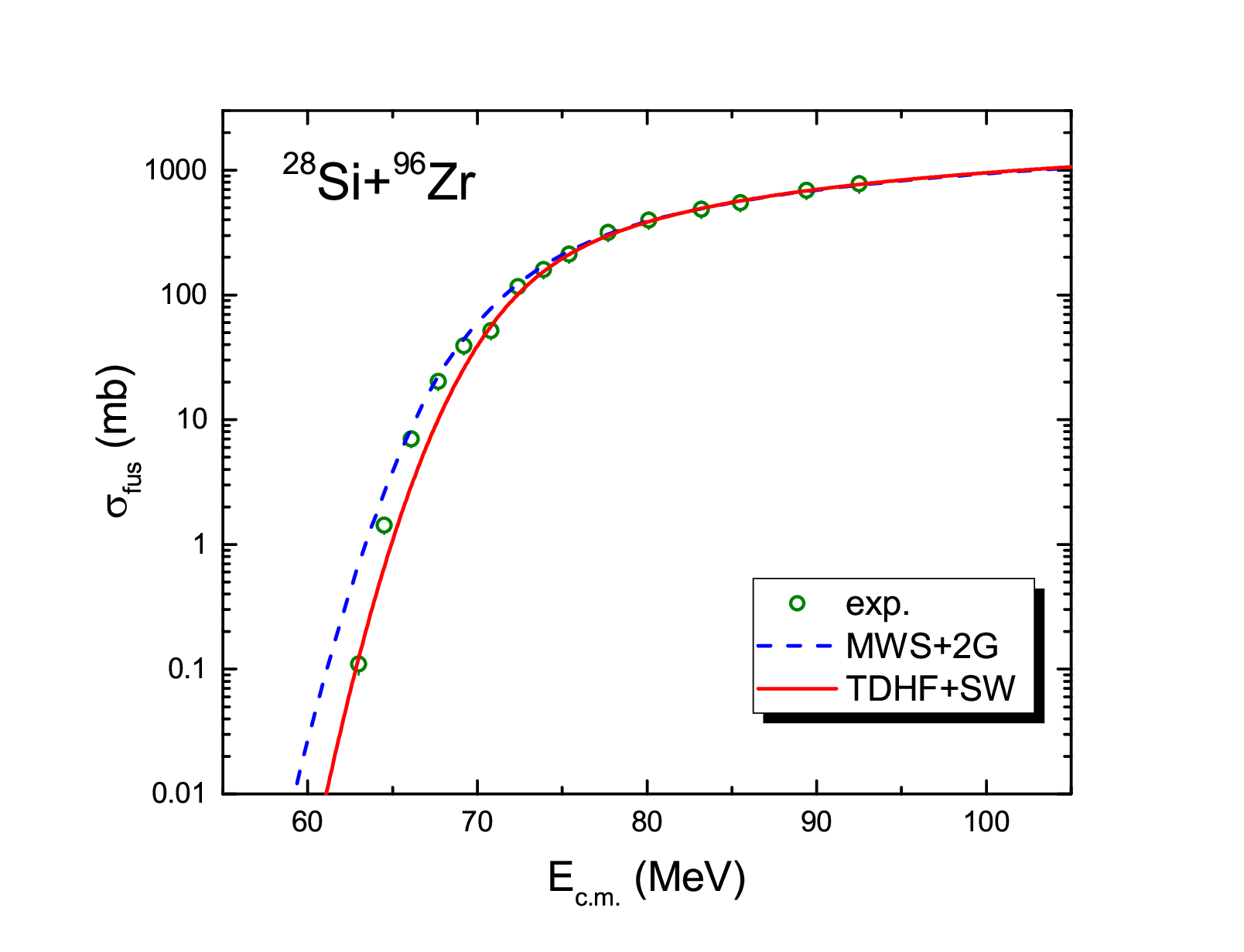}
        \caption{(Color online) The same as Fig. 5, but for $^{28}$Si+$^{96}$Zr. The data are taken from \cite{Khu17}.}
    \end{figure}

    To test the validity of the approach for description of the reactions with deformed projectile bombarding on nearly spherical target, we calculate fusion cross sections of $^{28}$Si+$^{96}$Zr and the results are shown in Fig. 7. The obtained capture barrier is $V_B=70.55$ MeV according to Eq.(5) and the value of $W$ is 2.73 MeV. We get a value of $R_B=10.3$ fm for the barrier radius from the FHF calculations with the weights similar to Eq.(3). One can see that the experimental data can be reasonably well reproduced. We also note that the calculated cross sections using the CCFULL program with coupling of the two-neutron transfer channel in addition to inelastic excitations are significantly lower than the data at sub-barrier energies, and the authors conclude that multi-neutron transfer channels appear to be important for $^{96}$Zr \cite{Khu17}. In the TDHF calculations, not only the static and dynamical deformation effects, but also multi-neutron transfer due to the excitation energy at capture position are self-consistently involved at the one-body level.

    \begin{figure}
        \setlength{\abovecaptionskip}{-2.7cm}
        \includegraphics[angle=0,width=0.85\textwidth]{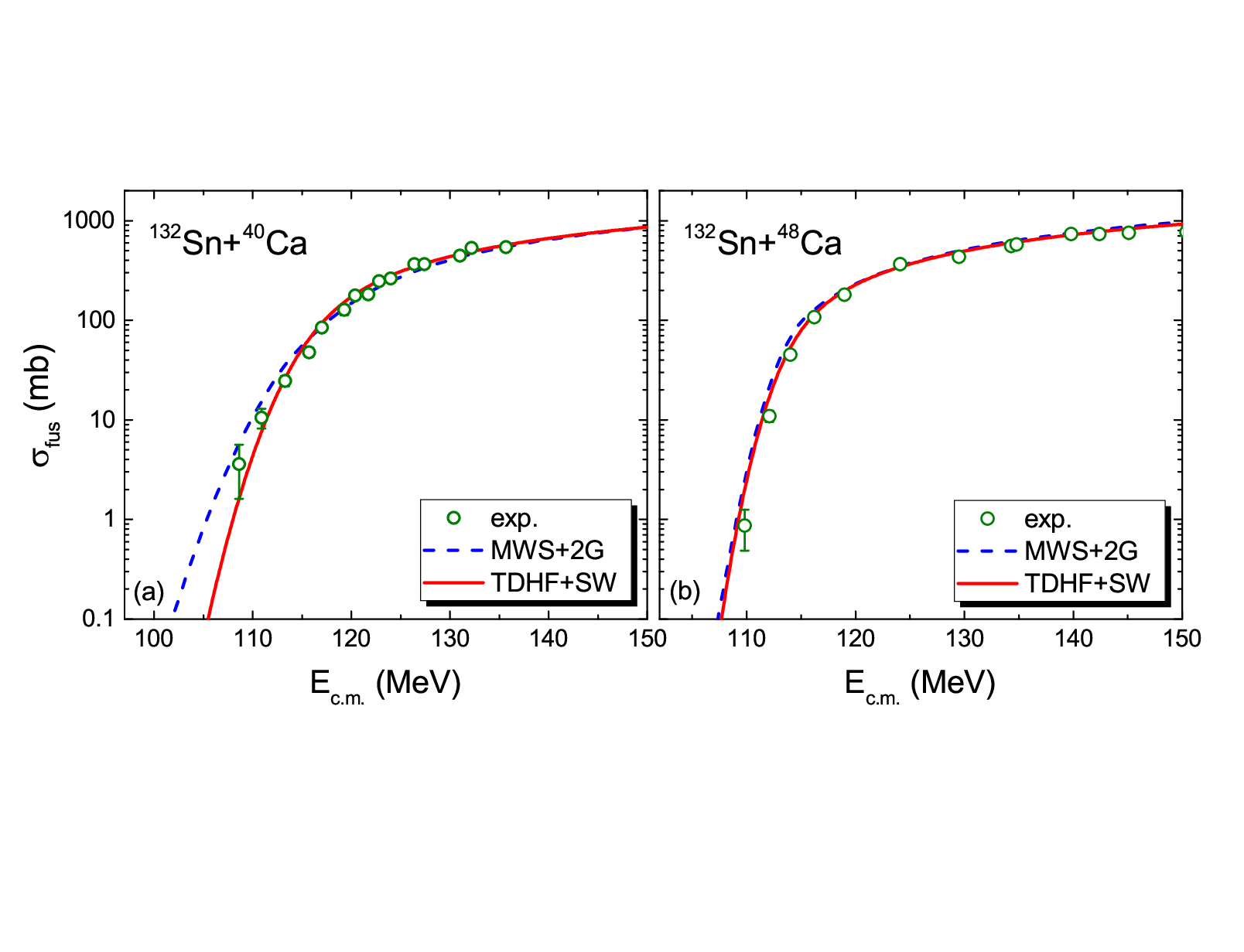}
        \caption{(Color online) The same as Fig. 5, but for $^{132}$Sn+$^{40,48}$Ca. The data are taken from \cite{Kola12}.}
    \end{figure}

    To further test the validity of the approach for description of the reactions with extremely neutron-rich nuclei, we study the fusion reactions $^{132}$Sn+$^{40,48}$Ca. The DC-TDHF potentials for these two reactions were studied in Ref. \cite{God17}, and the authors noted that the isovector reductions are quite different in these two reactions. Our calculated fusion cross sections are shown in Fig. 8. For $^{132}$Sn+$^{40}$Ca, the excitation energy of the compound nucleus is as high as about 62 MeV at an incident energy of $E_{\rm c.m.}=V_B$, which probably results in multi-nucleon transfer in the capture process and broadens the width of the barrier distribution. With the obtained barrier height $V_B=114.38$ MeV from the TDHF calculations and the value of $W=3.24$ MeV according to the reaction $Q$-value, the measured fusion excitation function can be reasonably well reproduced. Although with 8 neutrons more in $^{48}$Ca, the excitation energy of the compound nucleus in $^{132}$Sn+$^{48}$Ca is only 37 MeV, smaller than that of $^{132}$Sn+$^{40}$Ca by 25 MeV at $E_{\rm c.m.}=V_B$, due to the stronger shell effect in $^{48}$Ca. We note that the measured charge radius of $^{48}$Ca  is even slightly smaller than that of $^{40}$Ca \cite{Wang13} and the shell correction (in absolute value) for $^{48}$Ca is larger than that of $^{40}$Ca by 3.7 MeV according to the Weizs\"acker-Skyrme (WS4) mass model \cite{WS4} calculations. The relatively smaller excitation energy in $^{132}$Sn+$^{48}$Ca implies the dynamical deformation and nucleon transfer is more difficult in comparison with $^{40}$Ca, and leads to a much smaller value of $W=1.92$ MeV comparing with the value of $W=3.24$ MeV for $^{132}$Sn+$^{40}$Ca. From Fig. 8, one can see that the measured fusion cross section of $^{132}$Sn+$^{48}$Ca at $E_{\rm c.m.}\approx 110$ MeV is one order of magnitude lower than that of $^{132}$Sn+$^{40}$Ca, which can be remarkably well reproduced with the TDHF calculations. The competition between enhancement and suppression effects on sub-barrier fusion caused by excess neutron effects and neutron-shell-closure was also observed in \cite{liumin}.

    \begin{center}
        \textbf{IV. SUMMARY}
    \end{center}

    Based on the microscopic TDHF theory, we systematically investigate the capture thresholds $E_{\rm cap}$ for 144 fusion systems with nearly spherical nuclei. We find that the calculated $E_{\rm cap}$ for the reactions between doubly-magic nuclei are in good agreement with the extracted barrier heights from measured fusion excitation functions. For the fusion reactions with nearly spherical (open shell) nuclei which have relatively lower excitation threshold in comparison with magic nuclei, an excitation energy of about 1 MeV at the capture position seems to be required to better reproduce the data. The rms deviation between the extracted barrier heights and those from the TDHF calculations for the 144 fusion reactions is 1.43 MeV, which is smaller than the results of three empirical nuclear potentials: including the Bass potential and two Woods-Saxon potentials. It should be mentioned that the uncertainty of EDFs could result in a few percent fluctuation in the barrier height \cite{God22}. Together with Siwek-Wilczy\'{n}ski formula in which the three parameters are determined by the TDHF calculations, the fusion excitation functions are further investigated. The measured fusion cross sections can be reasonably well reproduced for seven fusion reactions $^{40}$Ca+$^{48}$Ca, $^{16}$O+$^{208}$Pb, $^{28}$Si+$^{96}$Zr, $^{40}$Ca+$^{90,96}$Zr and $^{132}$Sn+$^{40,48}$Ca. We find that the excitation energy of the system at the capture position which is related to the reaction $Q$-value can affect  the fusion cross sections at sub-barrier energies for reactions with nearly spherical nuclei. A relatively higher excitation energy at capture position could result in stronger effects for dynamical deformations and nucleon transfer in the capture process and broadens the width of the barrier distribution.

    \begin{center}
        \textbf{ACKNOWLEDGEMENTS}
    \end{center}
    This work was supported by National Natural Science Foundation of
    China (Nos. 12265006, U1867212 ), Guangxi Natural Science Foundation (No. 2017GXNSFGA198001), and Innovation Project of Guangxi Graduate Education (YCSW2022176). N. W. is grateful to L. Guo, H. B. Zhou and C. J. Lin for valuable discussions. The table of the capture barrier heights with the TDHF calculations is available from  http://www.imqmd.com/fusion/Ecap.txt

\end{document}